\documentclass[prb,amsmath,amssymb,twocolumn,showpacs,floatfix]{revtex4}
\usepackage{graphicx}
\usepackage{dcolumn}
\usepackage{bm}

\begin{document}

\preprint{APS/123-QED}
\title{Spin-glass state of vortices in YBa$_2$Cu$_3$O$_y$ 
and La$_{2-x}$Sr$_x$CuO$_4$ below the metal-to-insulator crossover}

\author{J.E.~Sonier,$^{1,2,*}$ F.D.~Callaghan,$^{1}$ Y.~Ando,$^{3}$ R.F.~Kiefl,$^{2,4}$ 
J.H.~Brewer,$^{2,4}$ C.V.~Kaiser,$^{1}$ V.~Pacradouni,$^{1}$ S.A. Sabok-Sayr,$^{1}$
X.F.~Sun,$^{3}$ S.~Komiya,$^{3}$ W.N.~Hardy,$^{2,4}$ D.A.~Bonn$^{2,4}$, and  R.~Liang$^{2,4}$}

\email{jsonier@sfu.ca}

\affiliation{$^1$Department of Physics, Simon Fraser University, Burnaby, British Columbia V5A 1S6, Canada \\
$^2$Canadian Institute for Advanced Research, 180 Dundas Street West, Toronto, Ontario M5G 1Z8, Canada \\
$^3$Central Research Institute of Electric Power Industry, Komae, Tokyo 201-8511, Japan\\
$^4$Department of Physics and Astronomy, University of British Columbia, Vancouver, British Columbia V6T 1Z1, Canada} 

\date{\today}

\begin{abstract}
Highly disordered
magnetism confined to individual weakly interacting vortices 
is detected by muon spin rotation ($\mu$SR) in two different families of 
high-transition-temperature ($T_c$) superconductors, but only in samples on 
the low-doping side of the low-temperature normal state metal-to-insulator 
crossover (MIC). The results support an extended quantum phase transition 
(QPT) theory of competing magnetic and superconducting orders that incorporates 
the coupling between CuO$_2$ planes. Contrary to what has been inferred 
from previous experiments, the static magnetism that coexists with 
superconductivity near the field-induced QPT is not ordered.
Our findings unravel the mystery of the MIC and establish that the normal
state of high-$T_c$ superconductors is ubiquitously governed by a
magnetic quantum critical point (QCP) in the superconducting phase.
\end{abstract}

\maketitle

\section{Introduction}

For two decades, arrival at a firm theory for high-$T_c$ superconductivity 
has been hindered by an incomplete characterization of the phase diagram for 
cuprate materials. In zero field, 
$\mu$SR,\cite{Kiefl:89,Weidinger:89,Niedermayer:98,Panago:02,Kanigel:02,Sanna:04} 
NMR/NQR \cite{Julien:99} and neutron scattering 
\cite{Wakimoto:01,Stock:06} studies show that static (or quasistatic)
magnetism coexists with superconductivity in the underdoped regime. 
Field-induced or enhanced {\it static magnetic order} 
has also been clearly detected in underdoped La$_{2-x}$Sr$_x$CuO$_4$ 
(LSCO:$x$),\cite{Katano:00,Lake:02,Khaykovich:05} 
Pr$_{1-x}$LaCe$_x$CuO$_4$,\cite{Fujita:04,Kang:05} and La$_2$CuO$_{4+y}$ 
\cite{Khaykovich:02,Khaykovich:03}
by neutron scattering, and in underdoped Pr$_{2-x}$Ce$_x$CuO$_4$ 
(PCCO),\cite{Sonier:03} Pr$_{1-x}$LaCe$_x$CuO$_4$,\cite{Kadono:04} and 
LSCO:$x$\cite{Savici:05} by $\mu$SR.  
The neutron studies on LSCO:$x$ and La$_2$CuO$_{4+y}$ support a 
proposed phase diagram by Demler {\it et al.} \cite{Demler:01} 
in which the pure superconductor undergoes a QPT to a phase of coexisting
static magnetic and superconducting orders. A similar phase transition
compatible with the theory of Ref.~\cite{Demler:01} has also been 
observed in CeRhIn$_5$.\cite{Park:06}
Still the general applicability of this QPT model is questionable, 
since field-induced static magnetic order has not been established
in any of the other hole-doped cuprates. 
 
An important detail in the model of Ref.~\cite{Demler:01} is the
assumption that the vortices are two-dimensional (2D). In this case 
the competing order is stabilized only when there is strong overlap of
the 2D vortices within a CuO$_2$ layer. When this happens long-range 
magnetic order is established. However, Lake {\it et al.}
\cite{Lake:05} have shown that the field-induced order in LSCO:0.10
is in fact three-dimensional (3D), implying significant interlayer
coupling. Furthermore, a neutron/$\mu$SR study of LSCO:0.10 concluded 
that the vortices themselves are 3D.\cite{Divakar:04} Following the
work of Ref.~\cite{Demler:01}, Kivelson {\it et al.}
showed that competing order can be stabilized about a nearly isolated 
3D vortex.\cite{Kivelson:02}
The field-induced QPT in this extended 3D model is argued to be to a 
coexistence phase in which the spatial dependence of the competing 
order is substantially non-uniform.

Here we show that there is a generic field-induced 
transition to a coexistence phase where spin-glass-like (SG)
magnetism is confined to weakly interacting 3D vortices. 
The detection of this phase implies that the QPT previously identified 
in LSCO:$x$ by neutrons scattering is simply a crossover to 
a situation where competing static magnetism is
spatially uniform in the sample.
Furthermore, we identify the ``true'' field-induced QPT
as occurring near the critical doping for the low-temperature normal-state MIC
that occurs at a non-universal doping concentration in cuprate 
superconductors.
The insulating side of the normal-state MIC is characterized by a $\log(1/T)$
divergence of the in-plane resistivity $\rho_{ab}$,
\cite{Ando:95,Boebinger:96,Fournier:98,Ono:00,Li:02,Dagan:04,Vedeneev:04,Dagan:05}
but it has also been indirectly identified by electronic thermal conductivity
measurements.\cite{Sun:03,Hawthorn:03,Sun:04}

\section{Experiment}

Muon spin rotation/relaxation ($\mu$SR) measurements were performed 
at TRIUMF, Canada on LSCO:$x$ and YBa$_2$Cu$_3$O$_y$ (YBCO:$y$) single crystals on 
either side of the previously determined critical dopings $x_c \! \approx \! 0.16$ 
(Refs.~\cite{Boebinger:96,Sun:03,Hawthorn:03}) and $y_c \! \approx \! 6.55$ 
(Ref.~\cite{Sun:04}) for the low-temperature MIC. The LSCO:$x$ single crystals
were grown by the traveling-solvent floating-zone technique,\cite{Ando:01} whereas 
the YBCO:$y$ single crystals were grown by a self-flux method 
in fabricated BaZrO$_3$ crucibles.\cite{Liang:98}

The $\mu$SR method involves the implantation of nearly 100~\% spin polarized 
positive muons into the sample. Like a tiny bar magnet, the magnetic moment of
the muon precesses about the local magnetic field $B$ with an angular 
frequency $\omega_\mu \! = \! \gamma_\mu B$, where 
$\gamma_\mu \! = \! 0.0852$~$\mu$s$^{-1}$~G$^{-1}$ is the muon gyromagnetic ratio. 
By measuring the time evolution of the polarization of the ensemble of muon spins
$P(t)$ via the anisotropic distribution of decay positrons, the internal magnetic field 
distribution $n(B)$ of the sample is determined.\cite{Sonier:00} As described below, 
$\mu$SR measurements were first carried out in zero external field to search for
{\it static} electronic moments. The vortex cores were then probed by 
transverse-field (TF) $\mu$SR, with the applied magnetic field     
perpendicular to the CuO$_2$ layers.

\section{Zero-Field Measurements}

\begin{figure}
\centering
\includegraphics[width=10.0cm]{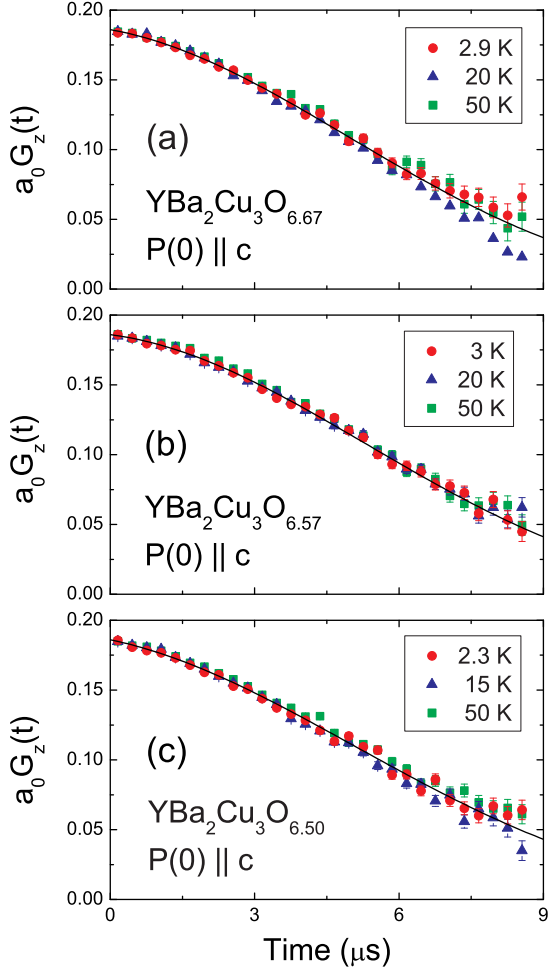}
\caption{(Color online) ZF-$\mu$SR time spectra at different temperatures
for (a) YBCO:6.67, (b) YBCO:6.57 and (c) YBCO:6.50. In all cases
the initial muon spin polarization ${\bf P}(0)$ was parallel
to the $\hat{c}$-axis of the single crystals. The solid curves
through the data points are fits described in the main text.}
\label{fig1}
\end{figure}

\begin{figure}
\centering
\includegraphics[width=10.0cm]{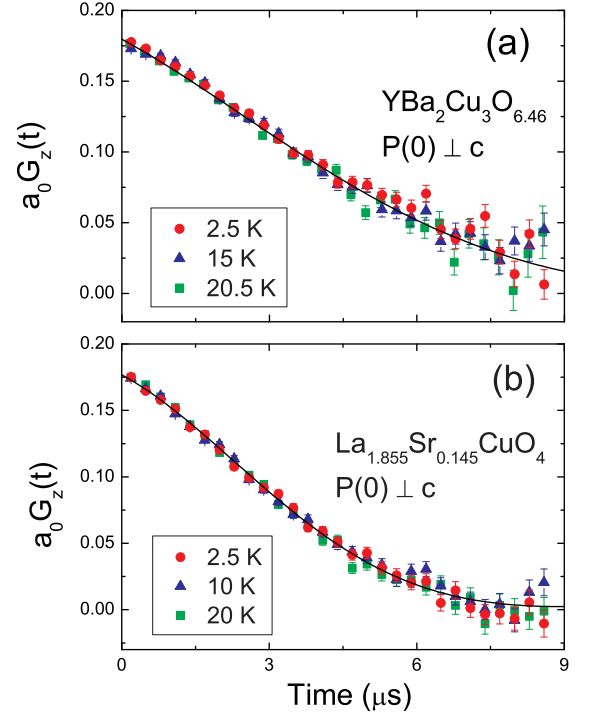}
\caption{(Color online) ZF-$\mu$SR time spectra at different temperatures
for the lowest doped samples (a) YBCO:6.46, and (b) LSCO:0.145, recorded with
the initial muon spin polarization ${\bf P}(0)$ perpendicular
to the $\hat{c}$-axis of the single crystals. The solid curves
through the data points are fits described in the main text.}
\label{fig2}
\end{figure}

\begin{figure*}
\centering
\includegraphics[width=19.0cm]{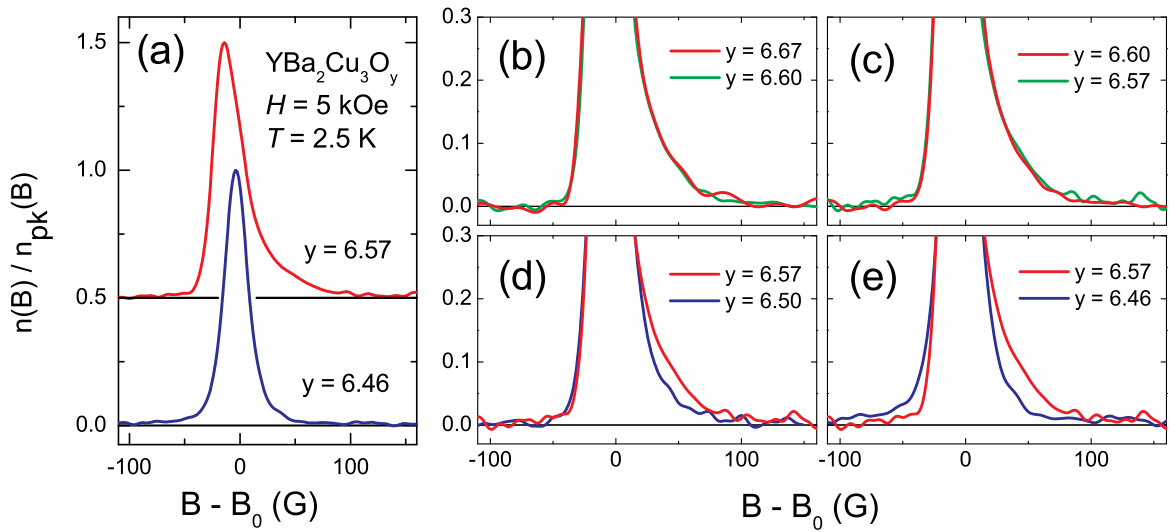}
\caption{(Color online) Doping dependence of the $\mu$SR line shapes for 
YBa$_2$Cu$_3$O$_y$ at $H \! = \! 5$~kOe and $T \! = \! 2.5$~K. 
(a) Full $\mu$SR line shape for samples above 
($y \! = \! 6.57$) and below ($y \! = \! 6.46$) the critical doping $y_c \! = \! 6.55$
for the MIC. (b), (c), (d), (e) Blowups of the `tail' region of the $\mu$SR line shapes 
above ($y \! = \! 6.67$, 6.60 and 6.57) and below ($y \! = \! 6.50$ 
and 6.46) $y_c \! = \! 6.55$. For comparison, all line shapes have been 
normalized as described in the main text.} 
\label{fig3}
\end{figure*}

Figures~\ref{fig1} and \ref{fig2} show ZF-$\mu$SR time spectra for some of 
the samples. Defining the direction of the initial muon spin polarization 
${\bf P}(0)$ to be parallel to the $\hat{z}$-axis, the ZF-$\mu$SR or 
``asymmetry'' spectrum has the form
\begin{equation}
A(t) = a_0 P_z(t) = a_0 G_z(t) \, ,
\end{equation}
where $a_0$ is the initial asymmetry and $G_z(t)$ is a relaxation
function. In all cases the spectra are well described
by the following ZF relaxation function   
\begin{equation}
G_z(t) =  G_z^{\rm KT}(t) \exp(- \lambda t) \, ,
\label{eq:relax}
\end{equation}
where
\begin{equation}
G_z^{\rm KT}(t) = \frac{1}{3} + \frac{2}{3}(1-\Delta^2 t^2) 
\exp \left( -\frac{1}{2}\Delta^2 t^2 \right) \, ,
\end{equation}
is the static Gaussian Kubo-Toyabe function (KT)
typically used to describe relaxation due to nuclear dipole fields, 
and $\lambda$ is an additional exponential relaxation rate. 
In the absence of static or slowly fluctuating electronic moments, 
the relaxation of the ZF-$\mu$SR signal is caused solely by the 
nuclear dipoles. In this case the relaxation is expected to be 
independent of temperature, as observed in Figs.~\ref{fig1} and 
\ref{fig2}. 
Fitted values for $\Delta$ and $\lambda$ are given
in Table~\ref{ZFresults}. The measurements on YBCO:6.46 and LSCO:0.145 
were done using a different spectrometer, and with the initial muon
spin polarization ${\bf P}(0)$ perpendicular, rather than parallel
to the $\hat{c}$-axis. In this geometry the relaxation rate is larger
due to the anisotropy of the muon-nuclear dipole interaction.\cite{Kiefl:90} 
The hole doping dependence of $\Delta$
in YBCO:$y$ is explained by a change in the ratio of muons stopping near the 
O(1) and O(4) oxygen sites.\cite{Sonier:02}       
While there is a residual exponential relaxation rate for 
all samples, $\lambda$ is independent of both temperature and 
hole doping concentration. Thus there is no evidence from the ZF-$\mu$SR
spectra for static electronic moments in our samples, which is
an essential requirement for establishing the presence of hidden 
competing magnetic order. We remark that the temperature-independent 
exponential component may come from the fraction of muons missing the 
sample and avoiding the background suppression scheme of the spectrometer.
Furthermore, the measurements here do not rule out the 
presence of a weak temperature dependent relaxation rate found in
earlier high precision ZF-$\mu$SR measurements of YBCO:$y$.\cite{Sonier:02,Sonier:01}       

\begin{table}
\caption[Paramters]{Results of fits of the ZF-$\mu$SR time spectra 
to Eq.~(\ref{eq:relax}). The fits are shown as solid curves in 
Figs.~\ref{fig1} and \ref{fig2}.} 
\begin{center}
\begin{tabular}{l l l l}
\hline \hline 
Sample & $\Delta$ ($\mu$s$^{-1}$) & $\lambda$ ($\mu$s$^{-1}$) & Polarization \\
\hline
YBCO:6.50 & 0.1120(3) & 0.0381 & ${\bf P}(0) \parallel \hat{c}$ \\
YBCO:6.57 & 0.1194(2) & 0.0244 & ${\bf P}(0) \parallel \hat{c}$ \\
YBCO:6.67 & 0.1207(1) & 0.0332 & ${\bf P}(0) \parallel \hat{c}$ \\
                                                                \\
YBCO:6.46 & 0.1277 & 0.1044 & ${\bf P}(0) \perp \hat{c}$ \\
LSCO:0.145 & 0.195(10) & 0.11(2) & ${\bf P}(0) \perp \hat{c}$ \\
\hline \hline
\end{tabular}
\label{ZFresults} 
\end{center}
\end{table} 

\section{Transverse-Field Measurements}

\begin{figure*}
\centering
\includegraphics[width=19.0cm]{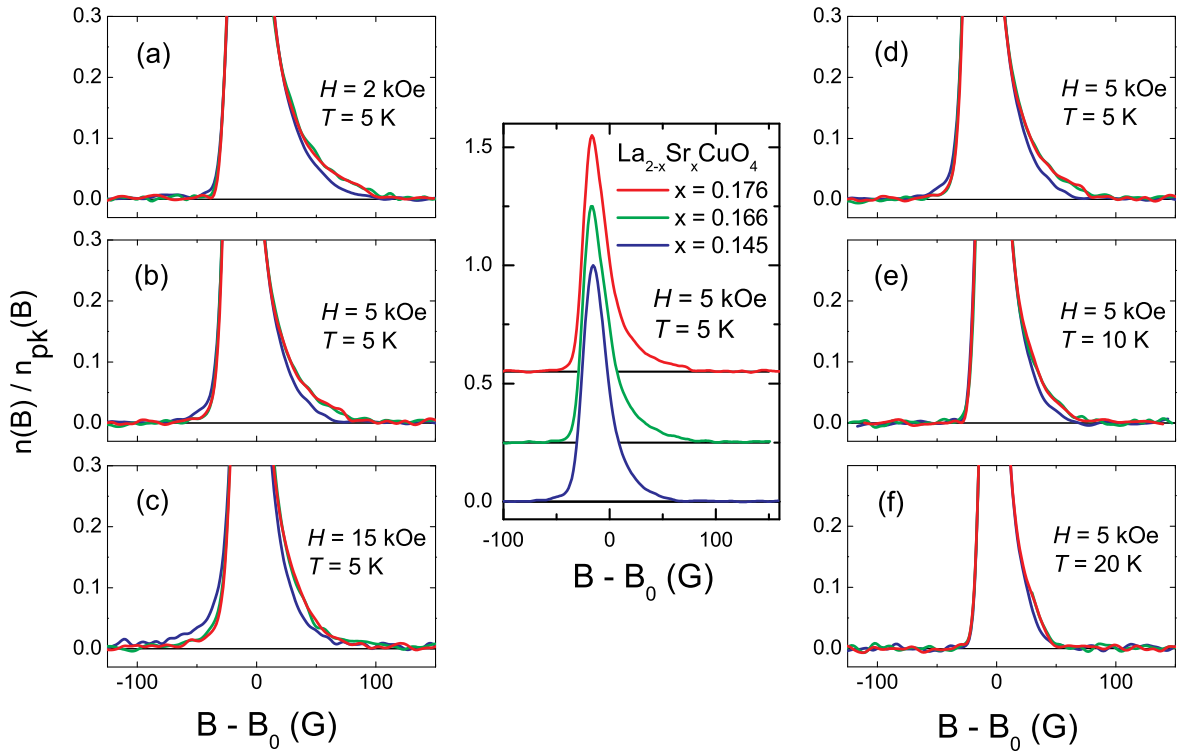}
\caption{(Color online) Doping, temperature and magnetic field dependences 
of the $\mu$SR line shapes for La$_{2-x}$Sr$_x$CuO$_4$.
The center panel shows the full $\mu$SR line shapes at $H \! = \! 5$~kOe
and $T \! = \! 5$~K, above ($x \! = \! 0.176$ and $x \! = \! 0.166$) and 
below ($x \! = \! 0.145$) the critical doping $x_c \! = \! 0.16$ 
for the MIC. Panels (a)-(c) show the field dependence of the `tail' regions 
of the normalized $\mu$SR line shapes. Panels (d)-(f) show the temperature 
dependence of the `tail' regions of the normalized $\mu$SR line shapes.}
\label{fig4}
\end{figure*}

In a transverse field, 
the muon spin precesses in a plane perpendicular
to the field axis. In this case the asymmetry spectrum is
\begin{equation}
A(t) = a_0 P_x(t) = a_0 G_x(t) \cos(\gamma_\mu B t) \, ,
\end{equation}
where $G_x(t)$ is the transverse muon spin relaxation function
and $B$ is the local field at the muon site.
In the vortex state the internal magnetic field is spatially
inhomogeneous, and the TF-$\mu$SR signal for a perfectly 
ordered flux-line lattice (FLL)
is described by the polarization function
\begin{equation}
P_x(t) = \sum_i \cos[\gamma_\mu B(r_i) t] \, ,  
\end{equation}
where the sum is over all sites in the real-space unit cell of the 
FLL and $B(r_i)$ is the local field at 
position $r_i \! = \! (x_i, y _i)$.
A Fourier transform of $P_x(t)$ 
\begin{equation}
n(B) = \int_{0}^{\infty} P_x(t) e^{-i (\gamma_\mu B  t)} e^{- \sigma_{\rm A}^2 t^2/2} dB \, ,
\end{equation}
often called the `$\mu$SR line shape', provides a fairly accurate visual 
illustration of the internal magnetic field distribution sensed by the muons.
Here $\exp(- \sigma_{\rm A}^2 t^2/2)$ is a Gaussian apodization function
used to suppress the ``ringing'' effect of the finite time range of $P_x(t)$. 
Figure~\ref{fig3}(a) shows a couple of examples of the $\mu$SR line shape
for YBCO:$y$ at $H \! = \! 5$~kOe and $T \! = \! 2.5$~K. The asymmetric
line shape for YBCO:6.57 is typical of the field distribution for
a 3D FLL.\cite{Sonier:00} Specifically, the `high-field' tail corresponds
to the spatial region in and around the vortex cores. 

Figures~\ref{fig3}(b)-(e)
show blowups of the `tail' regions of the Fourier transforms of TF-$\mu$SR  
spectra measured in the vortex state of YBCO:$y$ near the critical 
doping $y_c \! \approx \! 6.55$. For comparison, the line shapes have been 
normalized to their respective peak amplitude $n_{\rm pk}(B)$. Furthermore, 
to account for differences in the in-plane magnetic penetration depth, 
the widths of the line shapes have been made equivalent by rescaling the horizontal 
$B \! - \! B_0$ axis, where $B_0$ is the applied magnetic field.
Above $y_c$ the $\mu$SR line shapes for $y \! = \! 6.67$, 
$y \! = \! 6.60$ and $y \! = \! 6.57$ are identical. However, 
at $y \! = \! 6.50$ there is a clear {\it suppression} 
of the high-field tail, corresponding to the spatial region of the vortex cores.
Note that in a previous high-field study of YBCO:6.50 the high-field tail was 
argued to be enhanced rather than suppressed.\cite{Miller:02}
However, this conclusion was based on a comparison of the $\mu$SR line shape 
to an assumed theoretical curve for $n(B)$. As observed in Fig.~\ref{fig3}(e),
at $y \! = \! 6.46$ the suppression of the high-field tail
is accompanied by the appearance of a low-field tail.   
 
As shown in Fig.~\ref{fig4}, similar differences are observed
between the $\mu$SR line shapes of LSCO:$x$ above and below the critical 
doping $x_c \! = \! 0.16$ for the MIC. With increasing magnetic field the  
differences between the tails of the line shapes are enhanced [see
Figs~\ref{fig4}(a), \ref{fig4}(b) and \ref{fig4}(c)]. On the other hand,
with increasing temperature the $\mu$SR
line shape of LSCO:0.145 becomes more like that of the samples 
above $x_c \! = \! 0.16$ [see Figs.~\ref{fig4}(d), \ref{fig4}(e) and \ref{fig4}(f)].
In the next section we explain how field-induced static electronic moments
in the samples on the low-doping side of the MIC
accounts for both the suppression of the high-field tail and the appearance 
of a low-field tail.

\section{Data Analysis}

While the change in the local magnetic field distribution in the region of the
vortex cores is evident from a visual inspection of the
$\mu$SR line shapes in Figs.~\ref{fig3} and \ref{fig4}, 
it is constructive to consider a simple analysis of the TF-$\mu$SR
time spectra. Recently we carried out a comprehensive analysis
of the $\mu$SR line shape in $y \! \geq \! 6.57$ single crystals.\cite{Sonier:07}
There we showed that the TF-$\mu$SR signal is well described by 
the polarization function
\begin{equation}
P_x(t) = e^{-\sigma_{\rm eff}^2 t^2 /2} \sum_i \cos[\gamma_\mu B(r_i) t] \, ,  
\end{equation}
where the Gaussian function $\exp(-\sigma_{\rm eff}^2 t^2 /2)$ accounts for
additional relaxation due to FLL disorder and nuclear dipole moments,
the sum is over all sites in an hexagonal FLL, and $B(r_i)$ is the following
analytical solution of the Ginzburg-Landau equations\cite{Yaouanc:97} 
\begin{equation}
B(r_i) = B_0 \sum_{ {\bf G}}
\frac{e^{-i {\bf G} \cdot {\bf r}_i} \, \, F(G)}{\lambda_{ab}^2 G^2} \, .
\label{eq:GLfield}
\end{equation}
Here {\bf G} are the reciprocal lattice vectors of the FLL, 
$B_0$ is the average internal magnetic field,
$F(G) = u K_1(u)$ is a cutoff function for the {\bf G} sum, 
$K_1(u)$ is a modified Bessel function, and $u = \sqrt{2} \xi_{ab} G$.
The cutoff function $F(G)$ depends on the spatial profile of the 
superconducting order parameter at the center of the vortex core.
Consequently, the parameter $\xi_{ab}$ is a measure of the vortex core size.
As explained in Ref.~\cite{Sonier:07}, only the $H \! \rightarrow \! 0$
extrapolated value of $\lambda_{ab}$ is a true measure of the
magnetic penetration depth, since at finite $H$ this parameter absorbs
deviations of $B(r_i)$ from Eq.~(\ref{eq:GLfield}). 

In Ref.~\cite{Miller:02} it was assumed that 
the unusual $\mu$SR line shape of YBCO:6.50 results from static 
antiferromagnetic order 
in the vortex cores. However, field-induced static magnetic order 
has never been observed in YBCO:$y$ by neutron scattering. 
Furthermore, Khaykovich {\it et al.} have shown by neutron 
scattering that static magnetic order occurs in LSCO:0.144 only above 
$H \! \approx \! 30$~kOe.\cite{Khaykovich:05}
Thus the $\mu$SR line shapes of YBCO:6.46, YBCO:6.50 and LSCO:0.145 
presented here for $H \! \leq \! 15$~kOe cannot be explained by static 
magnetic order in and around the vortex cores. 
Instead we consider the possibility that the weak fields
considered here induce disordered static magnetism, which 
is not ruled out by the neutron scattering experiments. A polarization 
function that describes the case of disordered static electronic moments 
in and around the vortex cores is
\begin{equation}
P_x(t) = e^{-\sigma_{\rm eff}^2 t^2 /2} 
\sum_i \exp(-\Lambda e^{-(r_i/\xi_{ab})^2} t) \cos[\gamma_\mu B(r_i) t].\
\label{eq:disAF}  
\end{equation} 
This equation simply says that a muon stopping at position $r_i$ in the
FLL experiences a Lorentzian distribution of fields typical of a SG system
that results in an exponential decay of $P(t)$. Furthermore, the exponential 
relaxation rate $\Lambda$, and hence the width of the field distribution, is 
assumed to decrease with increased distance from the center of the vortex core.

\begin{figure}
\centering
\includegraphics[width=10.0cm]{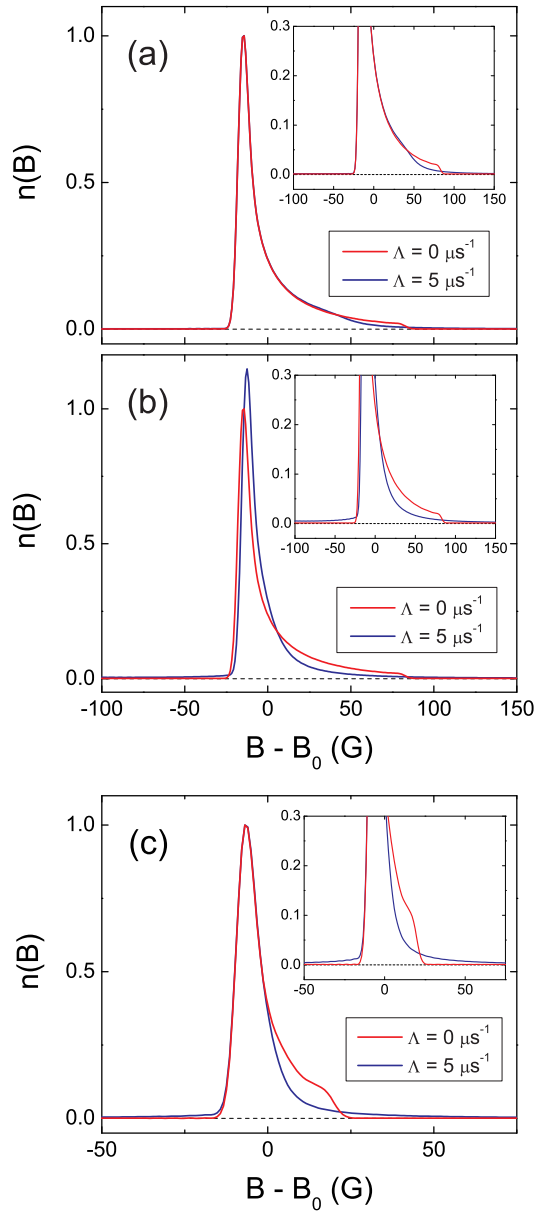}
\caption{(Color online) Real part of the Fourier transform of Eq.~(\ref{eq:disAF})
calculated for two values of $\Lambda$ (0~$\mu$s$^{ -1}$ and 5~$\mu$s$^{ -1}$),
with $\sigma_{\rm eff} \! = \! 0$ and 
(a) $\lambda_{ab} \! = \! 2000$~\AA, $\xi_{ab} \! = \! 50$~\AA, and
$B_0 \! = \! 5$~kG, (b) $\lambda_{ab} \! = \! 2000$~\AA, $\xi_{ab} \! = \! 100$~\AA, and
$B_0 \! = \! 5$~kG, and (c) $\lambda_{ab} \! = \! 3000$~\AA, $\xi_{ab} \! = \! 50$~\AA, 
and $B_0 \! = \! 15$~kG. All of the Fourier transforms have been generated with
a Gaussian apodization of width $\sigma_{\rm A} \! = \! 0.2$~$\mu$s$^{ -1}$.
The insets show blowups of the bottom portion of the same Fourier transforms.}
\label{fig5}
\end{figure}

Figure~\ref{fig5} shows the real part of the Fourier transform
\begin{equation}
n(B) = \int_{0}^{\infty} P_x(t) e^{-i \gamma_\mu B t} e^{- \sigma_{\rm A}^2 t^2/2} dt \, ,
\end{equation}
where $P_x(t)$ is calculated from Eq.~(\ref{eq:disAF}) for the case
$\sigma_{\rm eff} \! = \! 0$.
The vortex cores are non-magnetic for the case $\Lambda \! = \! 0$.
When $\Lambda$ is non-zero the high-field tail of the line shape is 
suppressed. With increasing $\Lambda$ the high-field tail is further
suppressed, and a low-field tail develops. While the change in the high-field 
tail is most recognizable, the appearance of the low-field tail depends on 
the width of the SG Lorentzian field distribution relative to 
the line width of $n(B)$ for the FLL. 
For example, in Figs.~\ref{fig5}(b) and \ref{fig5}(c) 
the Lorentzian field distribution of the 
static magnetism is broad enough to extend beyond the low-field cutoff of 
the field distribution of the FLL.    
  
\begin{figure}
\centering
\includegraphics[width=9.0cm]{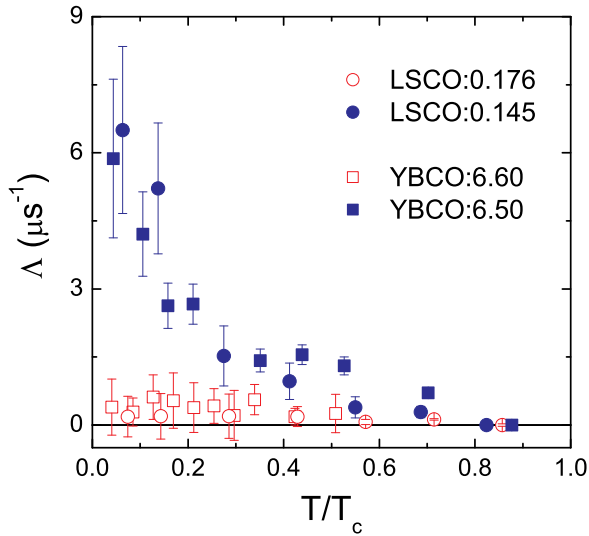}
\caption{(Color online) Temperature dependence of $\Lambda$ from fits of the TF-$\mu$SR 
time spectra at $H \! = \! 5$~kOe to Eq.~(\ref{eq:disAF}) 
for LSCO:$x$ and YBCO:$y$ samples above and below
the critical dopings $x_c \! = \! 0.16$ and $y_c \! = \! 6.55$, respectively.
At the lowest temperature the fits yield $\lambda_{ab} \! = \! 2505(30)$~\AA,
$\xi_{ab} \! = \! 34(3)$~\AA, and $\Lambda \! = \! 6.5(1.9)$~$\mu s^{-1}$
for LSCO:0.145, and $\lambda_{ab} \! = \! 1950(56)$~\AA,
$\xi_{ab} \! = \! 90(10)$~\AA, and $\Lambda \! = \! 5.9(1.8)$~$\mu s^{-1}$
for YBCO:6.50. Note that the vortex core size in YBCO:$y$ at low field
is large due to proximity-induced superconductivity 
on the CuO chain layers.\cite{Sonier:07}} 
\label{fig6}
\end{figure} 

Figure~\ref{fig6} shows the temperature dependence of $\Lambda$ obtained from
fits of the TF-$\mu$SR signal to Eq.~(\ref{eq:disAF})
for some of the samples. Since $\Lambda \! \approx \! 0$ for LSCO:0.176 
and YBCO:6.60, we conclude that the vortex cores are free of static 
magnetism. On the other hand, the diverging temperature dependence of $\Lambda$ 
for LSCO:0.145 and YBCO:6.50 indicates a static broadening of the internal magnetic
field distribution associated with the spatial region of the vortex cores.
In other words, the increase in $\Lambda$ with decreasing temperature is consistent
with a slowing down of fluctuating Cu spins. Moreover, since the value of $\Lambda$
does not saturate down to $T \! = \! 2.5$~K, the temperature dependence of
$\Lambda$ is consistent with an approach to a second-order magnetic phase 
transition at $T \! = \! 0$~K.

The main changes in the $\mu$SR line shape across the critical doping for the
low-temperature MIC are now understandable. From the fitted values of $\Lambda$,
the half-width at half-maximum of the Lorentzian field distribution assumed
in Eq.~(\ref{eq:disAF}) is approximately $\pm \! 70$~G in both LSCO:0.145 and 
YBCO:6.50 at $T \! = \! 2.5$~K. For LSCO:0.145 this is broad enough to affect
the low-field tail. With increasing field, the density of magnetic vortices 
increases, while the field inhomogeneity of the FLL decreases. 
Consequently, at higher magnetic field the static broadening of the $\mu$SR 
line shape by the magnetism becomes more discernable [see Figs.~\ref{fig4}(a)-(c)].
With increasing temperature, the simultaneous loss of 
the low-field tail and the recovery of the high-field tail of the LSCO:0145 line shape  
[see Figs.~\ref{fig4}(d)-(f)] signifies thermal destruction of the static
magnetism in and around the vortex cores.            

\section{Discussion}

\subsection{Dimensional crossover?}

While the onset of SG magnetism in and around the vortex cores fully
explains our experimental observations, we note that
the change in the $\mu$SR line shape across the MIC is somewhat reminiscent
of that observed across the 3D-to-2D vortex crossover field in highly anisotropic 
Bi$_{2+x}$Sr$_{2-x}$CaCu$_2$O$_{8+\delta}$ (BSCCO).\cite{Lee:93} In this case
random pinning-induced misalignment of the stacked 2D `pancake' vortices that
comprise the 3D flux lines in BSSCO, narrows and reduces the asymmetry of the 
$\mu$SR line shape. The two ingredients necessary for such a crossover are
weak coupling between the CuO$_2$ planes and a source of pinning.

If the vortices in LSCO:0.145 are quasi-2D, the gradual recovery of an asymmetric 
line shape at higher $T$ that is observed in Fig.~\ref{fig4} signifies thermal 
depinning of the vortices and a return to an ordered 3D FLL. Such a scenario
has been observed in BSCCO.\cite{Forgan:96} However, as already mentioned, 
the vortices in LSCO:$0.10$ are known to be 3D.\cite{Lake:05,Divakar:04} 
Since the effective mass anisotropy $\gamma \! = \! \sqrt{m_c^*/m_{ab}^*}$ increases
with decreasing hole doping concentration,\cite{Nakamura:93,Takenaka:94} 
a novel mechanism that 
softens the vortex lines at higher doping would be needed to explain the LSCO:0.145 
line shapes. As for YBCO:$y$, mutual inductance measurements show that even severely 
underdoped samples are quasi-2D only near $T_c$.\cite{Zuev:05}
The weak field dependence of the 
Josephson plasma resonance in YBCO:6.50 at low $T$ is also consistent with 3D 
vortices.\cite{Dulic:01} Thus the extreme vortex anisotropy necessary for
a 2D-to-3D crossover does not seem to occur at the hole-doping concentrations
investigated here. 

\begin{figure}
\centering
\includegraphics[width=10.0cm]{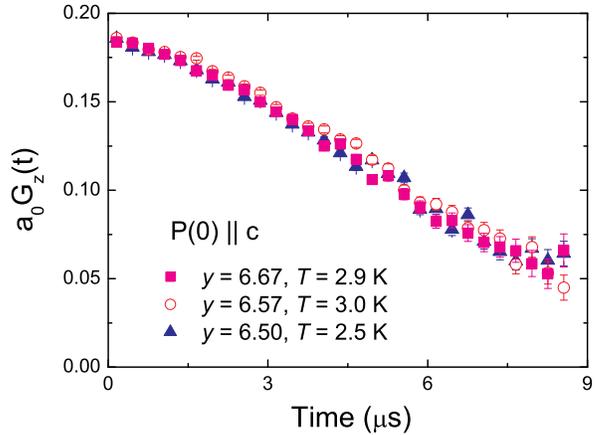}
\caption{(Color online) Low-temperature ZF-$\mu$SR time spectra for YBCO:$y$ 
on either side of the critical doping for the MIC at $y_c \! = \! 6.55$.}
\label{fig7}
\end{figure} 

\begin{figure*}
\centering
\includegraphics[width=26.0cm]{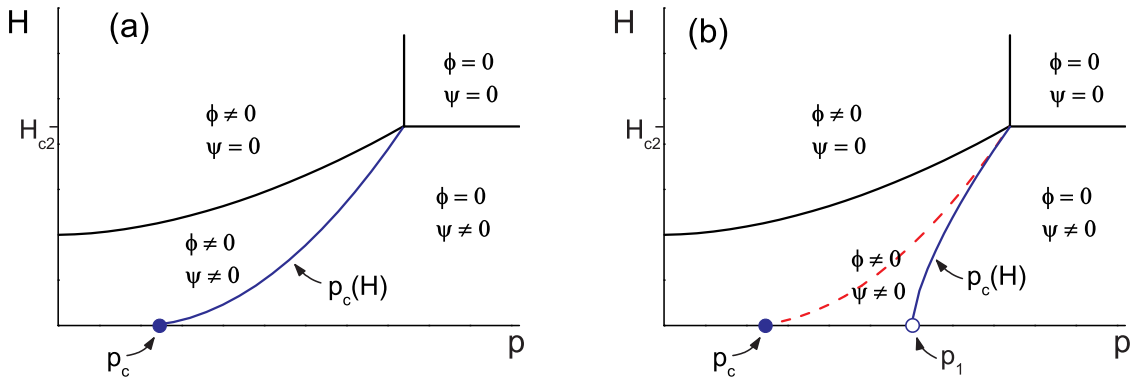}
\caption{(Color online) Schematic zero-temperature
$H$ versus hole doping concentration $p$ phase diagram for (a) 2D vortices 
(adapted from figure~1 of Ref.~\cite{Demler:01}), 
and (b) 3D vortices (adapted from figure~1 of Ref.~\cite{Kivelson:02}).
$\Psi$ and $\phi$ denote the expectation values of 
the superconducting and competing order parameters, respectively. 
In each panel the solid blue curve is a QPT and
the solid blue dot denoted $p_c$ is a QCP.
Note that the QPT in (a) becomes a crossover (red dashed curve)
in (b). Below the crossover, the competing order is spatially
non-uniform. In addition, the $H \! \rightarrow \! 0$ extrapolation of the 
QPT in (b) at $p_1$ is an `avoided' QCP (open blue circle). Note the
Meissner phase is not shown in (a) or (b).}
\label{fig8}
\end{figure*}

Assuming this were not the case, one could imagine an abrupt onset of disorder 
at the MIC that drives both the logarithmic divergence of the normal-state
resistivity and a 3D-to-2D vortex crossover. Since the location of the MIC in 
LSCO:$x$ has been independently confirmed,\cite{Boebinger:96,Sun:03,Hawthorn:03}
such pinning would have to be intrinsic to the material.
However, disorder in LSCO:$x$ decreases with decreasing Sr doping, and disorder due 
to excess or deficient oxygen primarily affects the lightly-doped and overdoped
regimes, respectively. Likewise, the onset of pinning below the MIC 
is inconsistent with ortho-II ordering in YBCO:6.50, which 
reduces random pinning by oxygen disorder and defects. An abrupt redistribution 
of charge at the MIC is also not supported by our own ZF-$\mu$SR measurements.
Previously, ZF-$\mu$SR studies of cuprate superconductors have demonstrated 
a sensitivity to charge-poor magnetic regions \cite{Niedermayer:98}, and to charge 
correlations \cite{Sonier:02}. However, as shown
in Fig.~\ref{fig7}, the low-temperature ZF-$\mu$SR signal does not change 
across the MIC.

\subsection{Avoided quantum criticality}
 
Figure~\ref{fig8} shows a comparison of the proposed phase diagrams 
for competing order in the cuprates for the case of 2D 
(Ref.~\cite{Demler:01}) and 3D (Ref.~\cite{Kivelson:02}) vortices.
The major difference is that the inclusion of the interlayer coupling 
allows the competing phase to be stabilized in nearly 
isolated vortices, thus altering the position and character of the QPT. 
There are two key predictions of the extended theory for 3D vortices
that are confirmed by our experiments. The first is that there
exists a coexistence phase of spatially inhomogeneous competing order.
At low fields where the interaction between vortex lines is weaker, 
we have detected SG magnetism that is localized in and 
around the vortex cores. This means that the competing order initially 
stabilized at the QPT is not static magnetic order as previously established,
but rather is highly disordered static magnetism. The competing
magnetism is characterized by a local order parameter, namely,
the mean squared local magnetization.                     
With increasing field, stronger overlap of the magnetism around 
neighboring vortices may lead to a co-operative 
bulk crossover to long-range magnetic order, as is apparently the case in 
La$_{1.856}$Sr$_{0.144}$CuO$_4$.\cite{Khaykovich:05}
While field-induced static magnetic order has not been detected in
YBCO:$y$, it is worth noting that the MIC occurs at a much lower hole doping 
concentration ($p_c \! \approx \! 0.1$) than in LSCO:$x$ ($p_c \! = \! 0.16$).
Consequently, very high magnetic fields are likely needed to induce
long-range magnetic order in YBCO:$y$. In contrast, very weak fields
were shown to induce magnetic order in PCCO samples that are below 
the MIC crossover at $p_c \! \approx \! 0.16$.\cite{Sonier:03} 
This is understandable, since in zero field the superconducting phase
of PCCO is in close proximity to the pure antiferromagnetic phase
where the competing order parameter is spatially uniform throughout
the sample.
 
Another key prediction of Ref.~\cite{Kivelson:02} is that there is 
an `avoided' QCP at $H \! = \! 0$, 
meaning that the QCP lies at a lower doping than the
extrapolated $H \! \rightarrow \! 0$ value of the field-induced QPT.
This is shown in Fig.~\ref{fig8}(b). 
Consistent with this idea, ZF-$\mu$SR studies of pure LSCO:$x$ 
\cite{Panago:02,Niedermayer:98} and
YBCO$y$ \cite{Kiefl:89,Sanna:04} indicate that the onset temperature 
for coexisting static magnetism and superconductivity
extrapolates to zero below the critical doping for the MIC.
While this is well below the doping concentration $p \! = \! 0.19$
that Tallon and others \cite{Tallon:01,Panago:02} have advocated
to be a universal QCP in the cuprates, we stress that
our study does not prohibit 
the existence of more than one QCP under the superconducting `dome'.
 
\section{Conclusions}

\begin{figure}
\centering
\includegraphics[width=9.0cm]{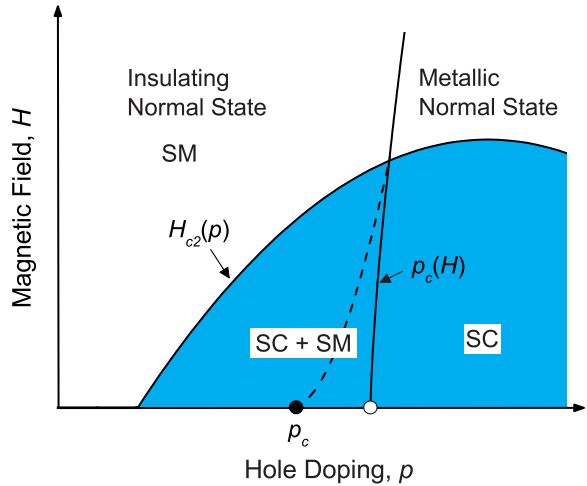}
\caption{(Color online) Schematic $T \! = \! 0$~K phase diagram deduced 
from this study. The normal and superconducting (SC) phases occur above and below 
the upper critical field $H_{c2}(p)$, respectively. The solid vertical curve at $p_c(H)$
is a QPT coinciding with the low-$T$ normal-state MIC. 
Below $H_{c2}(p)$, $p_c(H)$ separates a pure SC phase from a SC phase with coexisting 
static magnetism (SM). Immediately to the left of $p_c(H)$ the SM is 
disordered, becoming spatially uniform (and possibly ordered) above the dashed curve. 
The open circle is the predicted `avoided' QCP,\cite{Kivelson:02} 
whereas the solid circle indicates the `true' QCP at $H \! = \! 0$.}
\label{fig9}
\end{figure} 

Our experiments clearly demonstrate a change in the internal magnetic
field distribution of the vortex state across the critical doping 
concentration for the low-temperature MIC in two hole-doped high-$T_c$
superconductors. We have shown that the occurrence of SG magnetism
in and around weakly interacting vortices is the most likely source of 
the observed changes. In Fig.~\ref{fig9} we show a generic 
zero-temperature phase diagram that is compatible with the present 
and previous works. We conclude that the strange localization of
charge below the MIC stems from competing static magnetism that is
stabilized when superconductivity is suppressed by the applied
field. While others have hypothesized that magnetism is the cause
of the peculiar localization of charge, the experiments here
establish that static magnetism not present in zero external field
does appear in an applied magnetic field immediately below the 
critical doping for the MIC. Magnetism plays a prominent role 
in at least one theory for the MIC. In particular, Marchetti {\it et al.} 
have used a spin-charge gauge approach to show that the MIC can arise 
from a competition between
short-range magnetic order and the dissipative motion of the charge
carriers.\cite{Marchetti:01} The experiments here do not rule
out the possibility that there are short-range spin correlations 
in the field-induced magnetism immediately below the MIC.
          
We thank S.A. Kivelson, R. Greene, B. Lake and E. Demler for informative discussions. 
J.E. Sonier, J.H. Brewer, R.F. Kiefl, D.A. Bonn, W.N. Hardy and R. Liang acknowledge 
support from the Canadian Institute for Advanced Research and the Natural Sciences and 
Engineering Research Council of Canada. Y. Ando acknowledges support from 
Grant-in-Aid for Science provided by the Japan Society for the Promotion of Science.

\end{document}